\documentclass[12pt]{iopart}
\usepackage{iopams}
\usepackage{graphicx}
\usepackage{bm}
\usepackage{epstopdf}

\newcommand{\SrF}{SrFe$_2$As$_2$}
\newcommand{\CaF}{CaFe$_2$As$_2$}
\newcommand{\BaF}{BaFe$_2$As$_2$}
\newcommand{\LaCa}{Ca$_{1-x}$La$_x$Fe$_2$As$_2$}

\newcommand{\SrCax}{Sr$_{1-x}$Ca$_x$Fe$_2$As$_2$}
\newcommand{\SrCay}{Sr$_{0.3}$Ca$_{0.7}$Fe$_2$As$_2$}
\newcommand{\SrCaLa}{(Sr$_{1-y}$Ca$_y$)$_{1-x}$La$_x$Fe$_2$As$_2$}
\newcommand{\CaR}{Ca$_{1-x}$R$_x$Fe$_2$As$_2$}
\newcommand{\BaK}{Ba$_{1-x}$K$_x$Fe$_2$As$_2$}

\newcommand{\tc}{$T_c$}
\newcommand{\tn}{$T_N$}

\newcommand{\ie}{{\it i.e.}}

\begin{document}

\date{\today}

\title{Rare earth substitution in lattice-tuned \SrCay\ solid solutions}
%\title{Electron Doping Effects in \SrCa\ Solid Solutions}

\author{T.~ Drye, S.~R.~Saha, and Johnpierre~Paglione}

\address{Center for Nanophysics and
Advanced Materials, Department of Physics, University of Maryland,
College Park, MD 20742
}

\author{P. Y. Zavalij}
\address{Department of Chemistry and Biochemistry, University of
Maryland, College Park, MD 20742, USA}

\ead{paglione@umd.edu}

\begin{abstract}

The effects of aliovalent rare earth substitution on the physical properties of \SrCay ~solid solutions are explored.
Electrical transport, magnetic susceptibility and structural characterization data as a function of La substitution into \SrCaLa\ single crystals confirm the ability to suppress the antiferromagnetic ordering temperature from 200~K in the undoped compound down to $100$~K approaching the solubility limit of La. Despite up to $\sim 30\%$ La substitution, the persistence of magnetic order and lack of any signature of superconductivity above 10~K present a contrasting phase diagram to that of \LaCa, indicating that the suppression of magnetic order is necessary to induce the high-temperature superconducting phase observed in  \LaCa. 

\end{abstract}

\maketitle

\section{Introduction: }

The iron-pnictide superconductors have garnered much attention since the first observation~\cite{Kamihara}~that chemical manipulation of these materials could lead to high-temperature superconductivity \cite{Paglione}.  Particular interest has fallen on the structural ``122'' family with the ThCr$_2$Si$_2$ crystal structure, because of their intermetallic nature and relative ease of synthesis. Superconducting transition temperatures (\tc) as high as $25$ K have been observed in transition metal substituted \BaF~\cite{Nakajima}, while \tc ~up to $38$ K has been seen when hole doping on the alkaline earth site in \BaK \cite{Rotter}. 
Recently, superconductivity with \tc ~values reaching as high as $47$ K was observed to occur via aliovalent rare earth substitution into the alkaline earth site of \CaF~\cite{Saha1}, approaching the highest \tc ~values of all iron-based superconductors ($\sim55$~K in SmO$_{1-x}$F$_x$FeAs ~\cite{Ren1}). Stabilized by the suppression of antiferromagnetic order via a combination of both chemical pressure and electron doping, the superconductivity observed in the \CaR ~series (R = La, Ce, Pr, Nd) arises in a highly tunable system where the choice of rare earth species allows for structural tuning toward the collapsed tetragonal phase ~\cite{Hoffmann}, such as shown in the Pr- and Nd-doped series, or more simply electron doping with minimal change in unit cell size as shown for La substitution \cite{Saha1}.

The close ionic radius match of the light rare earths with Ca in the \CaF\ system makes for an ideal system with which to study the interplay between superconductivity and structural instability. However, the observations of partial volume-fraction diamagnetic screening in low-concentration rare earth-substituted \CaF\ \cite{Saha1,P Chu,Gao}, where the solubility limit of La is 30\%, poses a challenge to understanding the reasons for a lack of bulk-phase superconductivity. One promising route is through the inclusion of higher concentrations of rare earth substitution. Previous studies of La substitution into the \SrF ~system \cite{Muraba}, with \tc\ values up to ~$\sim22$~K, have shown a significant increase in the superconducting volume fraction when La content reaches $40$\% La for Sr, at which point the volume fraction jumps to nearly $70$\%. 
The introduction of higher concentrations of rare earth in the \CaF\ series is thus a promising route to achieving bulk superconductivity with high \tc\ values. 

While it has been shown that the application of pressure during synthesis can provide the desired results \cite{Muraba}, a clue to another possible route is provided by the materials themselves (\ie, in the case of the \LaCa\ series, where the $c$-axis remains constant but the $a$-axis increases with increasing rare-earth concentration up to the solubility limit of $\sim30\%$ \cite{Saha1}.)  It follows that a larger unit cell would likely be able to accomodate a larger concentration of rare earths.  It has also been shown previously that substituting Sr in the place of Ca leads to a controlled expansion of the unit cell in accordance with Vegard's Law \cite{Saha2}; therefore, such a solid solution may serve as the base formula into which rare earth atoms can be further doped.

In this article, we examine the feasibility of increasing the solubility limit of rare earth substitution by doping La into the (Sr,Ca)Fe$_2$As$_2$ solid solution system.  The synthesis of such single crystals with La substitution shows widely ranging chemical compositions and suggest a competition between Sr and La.  We detail the effects of increasing La content using systematic X-ray, electrical transport, and magnetization measurements, and compare these effects with those observed when La is doped into the \CaF ~parent material, specifically tracking the suppression of the AFM ordering temperature and signs of superconductivity in the system.  

%%%%Figure EDS Data
\begin{figure}[th]
\begin{center}
\includegraphics[width=16.0cm]{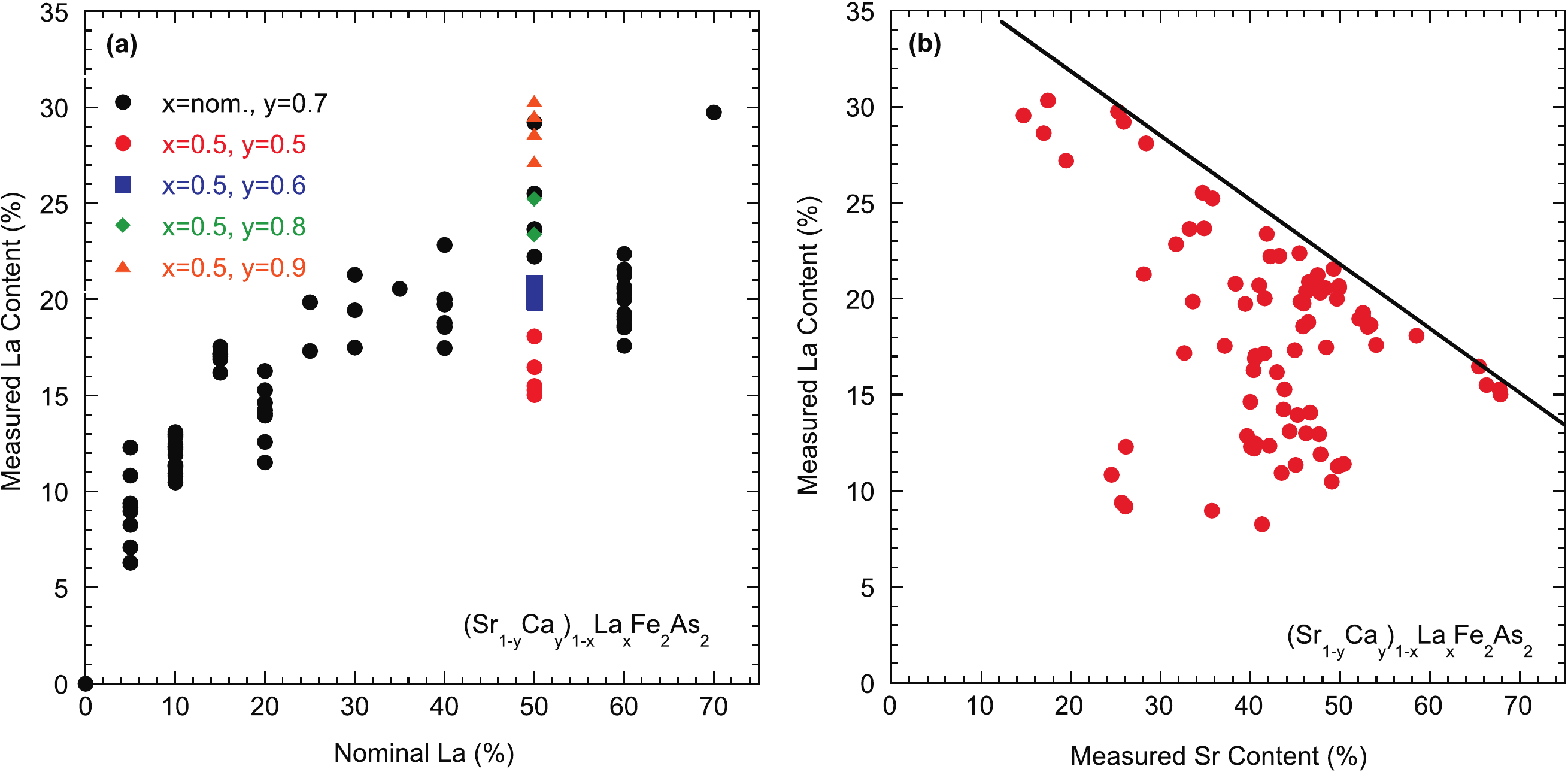}
\end{center}
\caption{\label{fig1}Results of energy dispersive spectroscopic (EDS) analysis of the elements occupying the alkaline earth site.  a)  Measured (actual) La content vs.\ Nominal (pre-reaction) La content.  Black close circles represent samples with pre-reaction stoichiometries of (Sr$_{0.3}$Ca$_{0.7}$)$_{1-x}$La$_x$Fe$_2$As$_2$.  Colored symbols represent samples with pre-reaction stoichiometries of (Sr$_{1-y}$Ca$_y$)$_{1-x}$La$_x$Fe$_2$As$_2$, where $y$ is denoted in the legend.  
b)  Measured La content vs.\ measured Sr content in the same samples shown in a).  All values are taken from EDS analysis.  The black solid line serves as the upper bound for the solubility of La.}
\end{figure}
%%%%%%%%%%%%%%

%%%%%%%%%%%%%%%%%%%%%%%%%%%%%%%%%%%%%%%%%
\section{Experiment}

Single crystals were grown via a self-f{}lux method using elemental stoichiometries of $(1-y)(1-x)$:$y(1-x)$:$x$:$4$ for Sr:Ca:La:FeAs according to the formula \SrCaLa ~with FeAs f{}lux~\cite{Saha3}.  Starting materials were placed inside alumina crucibles and sealed in quartz tubes under partial atmospheric pressure of Ar.  The growths were heated and allowed to slow cool, resulting in crystals with typical dimensions of ($5.0 \times 5.0 \times 0.10$) mm$^3$, which were mechanically separated from the frozen f{}lux. Chemical analysis was obtained using both energy-dispersive (EDS) and wavelength-dispersive (WDS) X-ray spectroscopy, showing $1$:$2$:$2$ stoichiometry between (Sr,Ca,La), Fe, and As concentrations.  EDS was conducted on a large number of samples in order to determine general concentration trends, while WDS was used to determine very accurately the concentrations of elements for samples used in X-ray, resistivity, and magnetization measurements.
Single crystal X-ray diffraction was performed on a Bruker Smart Apex2 diffractometer equipped with a CCD detector, graphite monochromator, and monocap collimator.  Crystal structures were refined (SHELXL-97 package) using the I4/mmm space group against 113 and 106 independent reflections measured at $250$ K and corrected for absorption using the integration method based on face indexing (SADABS software).  Because three different atoms were occupying the same crystallographic site, refinement of chemical compositions was not possible, and refinement instead was focused on obtaining unit cell parameters. Resistivity measurements were performed using the standard four-probe ac method, via gold wire and In/Sn solder contacts with typical contact resistance of $\sim0.5~\Omega$ at room temperature, using up to 1~mA excitation currents at low temperatures. Magnetic susceptibility was measured using a commercial superconducting quantum interference device magnetometer.

%%%%%%%%%%%%%%%%%%%%% RESULTS %%%%%%%%%%%%%%%%%%
\section{Results and Discussion}

%%%%Figure X-ray Data
\begin{figure}
\begin{center}
\includegraphics[width=16.0cm]{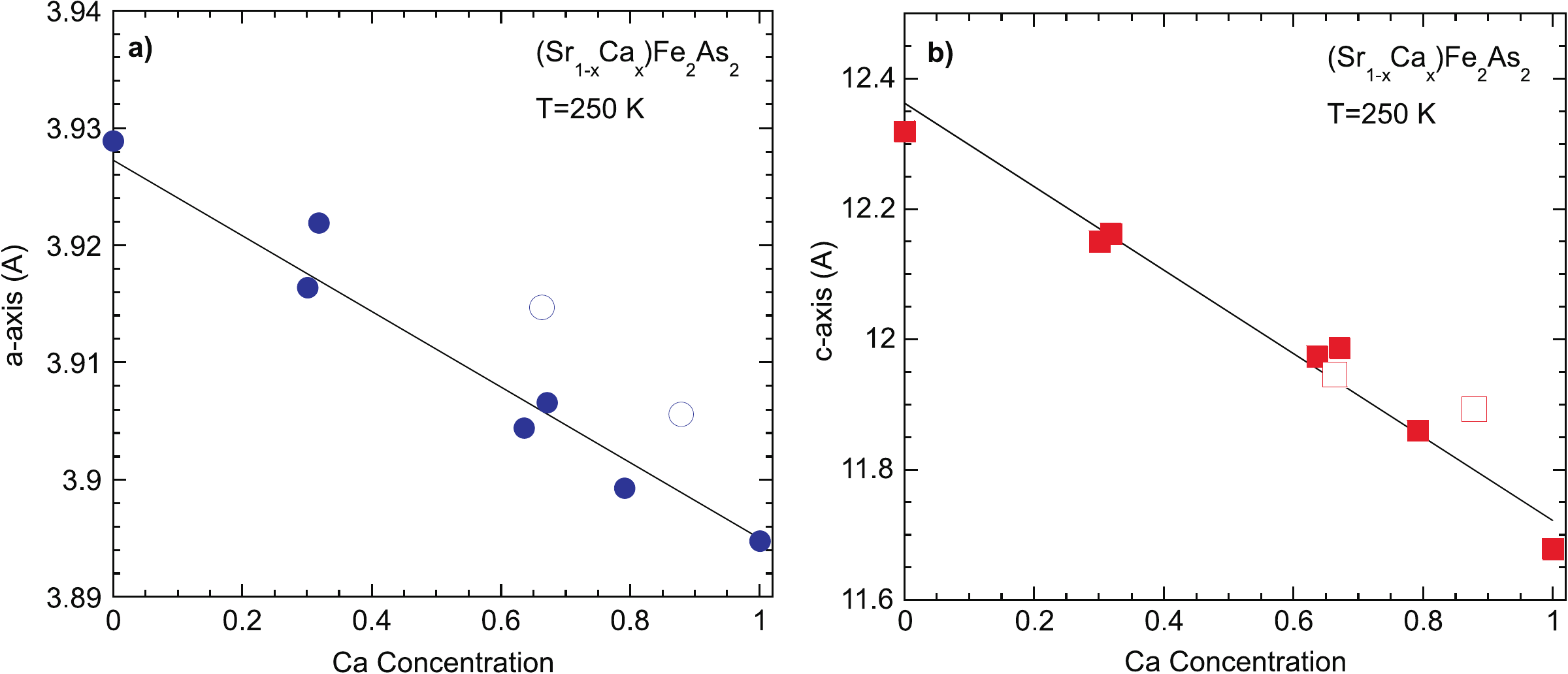}
\end{center}
\caption{\label{fig2}Unit cell parameters $a$ and $c$ obtained via single crystal x-ray diffraction of (Sr,Ca)Fe$_2$As$_2$ (filled points)~\cite{Saha2} and (Sr,Ca,La)Fe$_2$As$_2$ (open points), presented in (a) and (b), respectively. The $c$-axis lattice parameters are in good agreement, but the $a$-axis lattice parameters move away from the established trend as the amount of La increases.}
\end{figure}
%%%%%%%%%%%%%%

EDS of the actual concentration of La in the \SrCaLa ~series reveals that at low La values, the actual La content is higher than the nominal content, rising to a limit of  $\sim30$\% for growths with nominal La higher than $50$\% as shown in FIG. 1a.  While it is illustrated that increasing the starting ratio of Ca:Sr leads to higher La concentrations in the final materials, a stronger correlation between the Sr, Ca, and La concentrations can be found by plotting the measured La content against the measured Sr content as shown in FIG. 1b.  It is evident here that La and Sr are inversely correlated in this material and increasing the Sr concentration seems to strongly limit the amount of La that is able to dope into the sample. 

%%%%Figure Resistivity
\begin{figure}[tbh]
\begin{center}
\includegraphics[width=16.0cm]{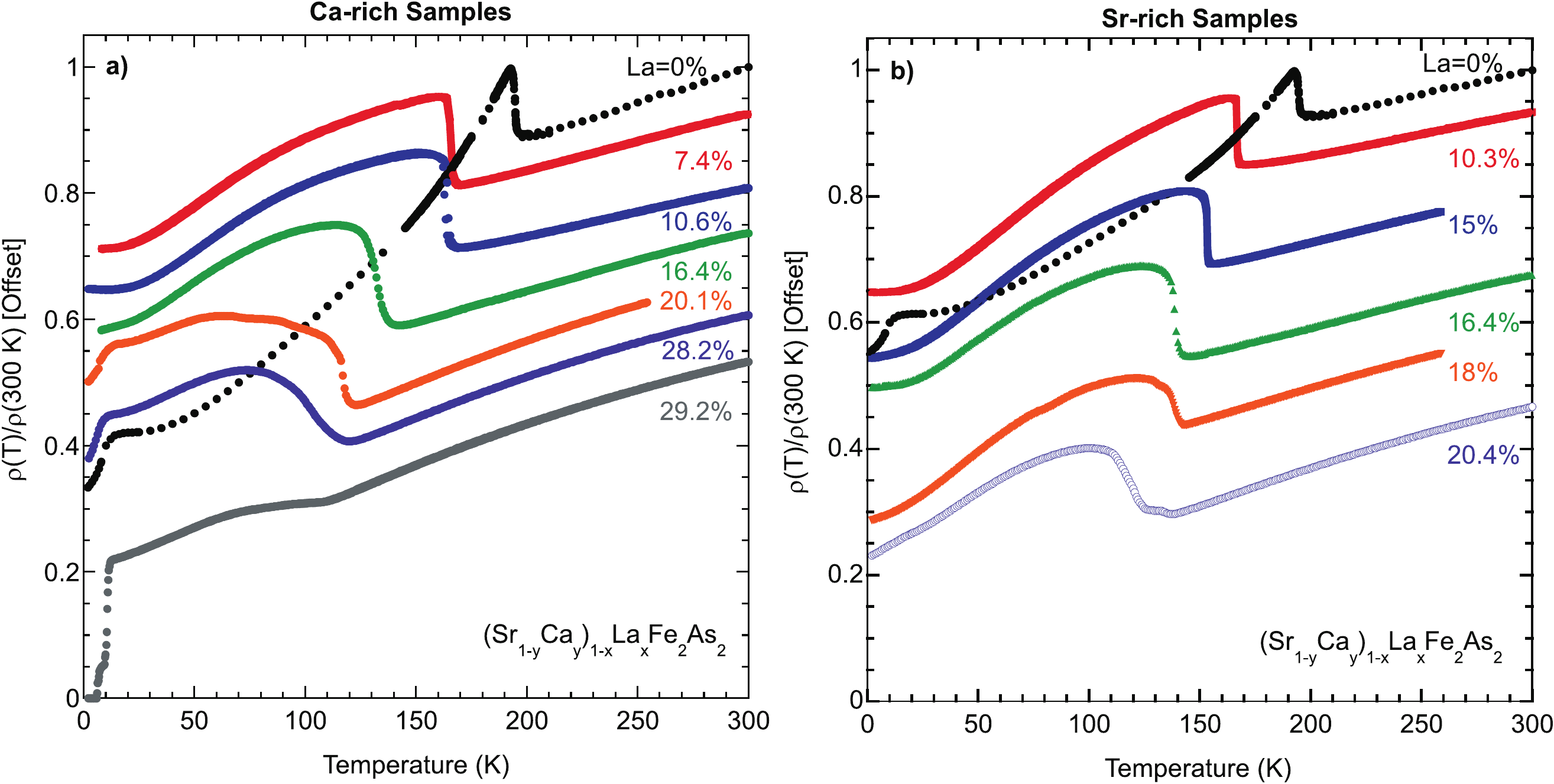}
\end{center}
\caption{\label{fig3}Resistivity of (Sr,Ca,La)Fe$_2$As$_2$ samples as a function of temperature, normalized to $300$ K and then vertically shifted for clarity; the Ca-rich samples are presented in (a), while the Sr-rich samples are presented in (b).}

\end{figure}
%%%%%%%%%%%%%%

Single crystal x-ray analysis allows us to analyze the progression of the lattice parameters as a function of the concentrations of Sr, Ca, and La in each sample.  In FIG. 2, we plot the lattice parameters of the samples used in this study alongside the lattice parameters observed for solid solutions of the parent compounds SrFe$_2$As$_2$ and CaFe$_2$As$_2$ ~\cite{Saha2}.
Previously, it was shown that doping La for Ca in \CaF ~results in a $c$-axis lattice parameter that does not change, despite expansion of the $a$-axis~\cite{Saha1}. Taking this into account, we have plotted these points as Sr$_{1-x}$(Ca,La)$_x$ -- the Ca and La values are taken together in order to determine the composition $x$, which places our points in good agreement with the $c$-axis values from the \SrCax ~study (FIG. 2b;, however, the $a$-axis values diverge as La content increases (FIG. 2a).  This implies that by selecting the proper Sr, Ca, and La content, we can tune the $a$-axis and $c$-axis parameters nearly independently.  This is in striking contrast to most doping studies on these materials, which show a strong coupling between $a$ and $c$-axes lattice parameters~\cite{Canfield}.

As seen in similar doping studies of iron-pnictides~\cite{Saha1,Muraba,Leithe}, it is expected that increasing the La content in these samples will be manifest in resistivity data as a systematic decrease in the Ne\'el ordering temperature \tn.  Electrical resistivity data of these samples (shown in FIG. 2) roughly resembles the expected behavior, as it is obvious that \tn ~is suppressed upon increased doping of La into the system.  A key difference here lies in the ranging chemical compositions obtained using WDS; subtraction of the La content leaves two classes of samples, \ie~the Ca-rich (shown in FIG. 3a) and the Sr-rich (shown in FIG. 3b). In the Sr-rich case, no sample was found to contain less than $\sim10$\% La or more than $\sim22$\% La, whereas in the Ca-rich case, a much wider range of La concentrations can be found (up to $\sim30$\% La). In the Sr-rich samples, \tn ~is gradually suppressed down to $\sim$130~K and no superconductivity is found to exist due to La substitution.  In the Ca-rich case, \tn ~is suppressed to a slightly lower value of $\sim$100~K, but no trace of a superconducting phase similar to that seen in the \LaCa~case~\cite{Saha1}, where high-\tc~values in the range 30-47~K are indicative of rare earth doping-induced superconductivity, was found. Note that the Ca-rich samples do exhibit traces of a superconductivity onset near $T^*\sim$10~K, which we attribute to the strain-induced phase often observed under non-hydrostatic pressure conditions \cite{Torikachvili,Kreyssig} and posited to nucleate at AFM domain walls~\cite{Xiao}. It is interesting to highlight the fact that this ``10 K'' phase  appears predominantly in Ca-rich samples, suggesting its stability is tied strictly to the \CaF\ magnetic and/or crystallographic structure.
  
%%%%Figure Magnetization
\begin{figure}[h]
\begin{center}
\includegraphics[width=10.0cm]{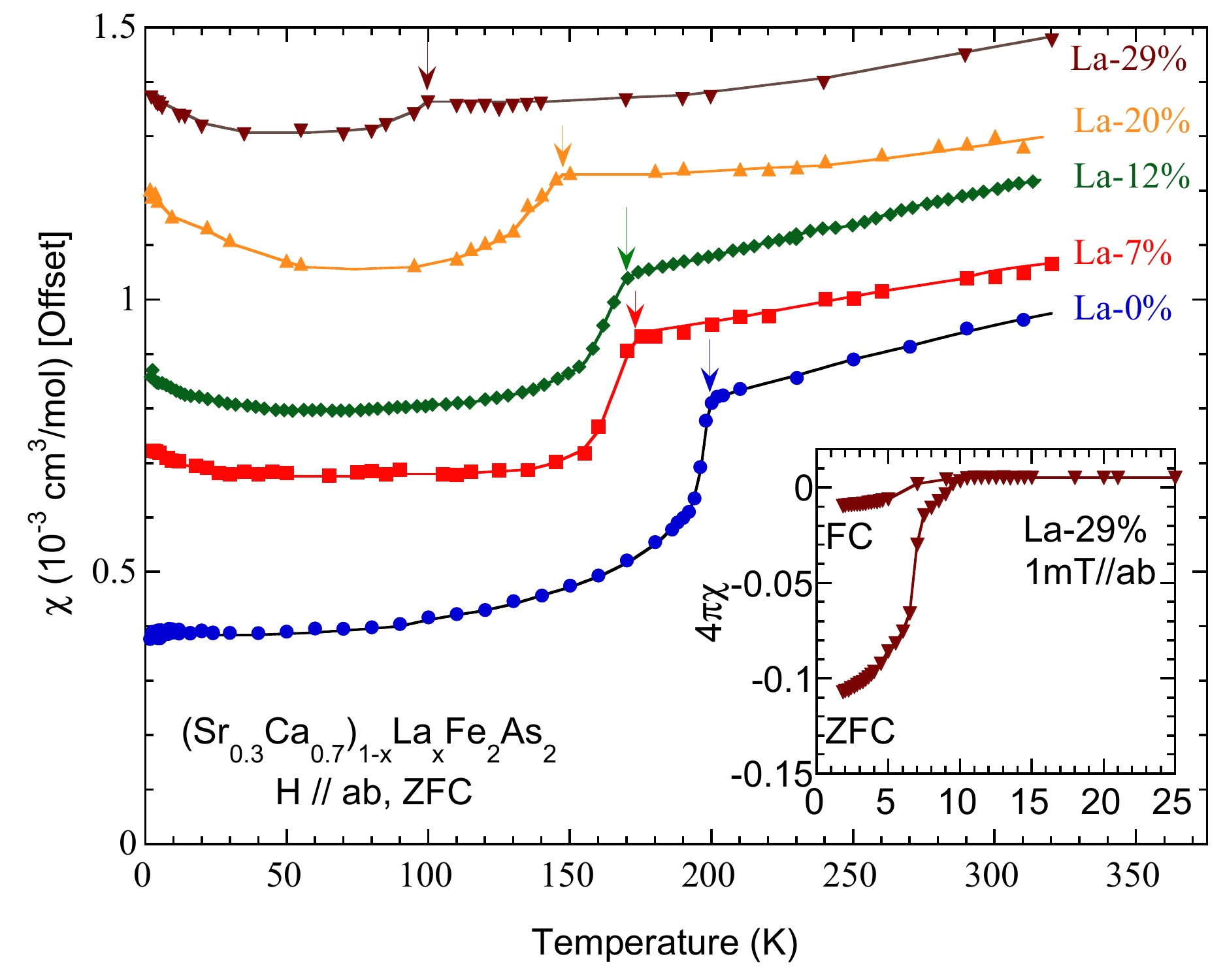}
\end{center}
\caption{\label{fig4}Magnetic susceptibility ($\chi$) vs.\ temperature for several La doped samples. The data are shifted along the vertical axis for clarity. The kink in each curve indicates the temperature of the antiferromagnetic transition, \tn~.  The inset shows the low field magnetic susceptibility of the $29$\%-La doped sample at low temperatures, where superconductivity below \tc~$\sim10$~K is seen; the estimated superconducting volume fraction is $\sim 11$\%.}
\end{figure}
%%%%%%%%%%%%%%

 Temperature dependence of magnetic susceptibility $\chi$($T$) data for Ca-rich samples (shown in FIG. 4) corroborates the picture drawn by electrical transport data. As expected from previous studies, \tn~is revealed as an antiferromagnetic ordering temperature traced by a kink in $\chi$($T$).  The suppression of \tn\ occurs at the same rate observed in transport data, with ordering at $\sim$100~K still present for samples which show superconductivity at $T^*\sim$10~K (inset of FIG. 4) at low field. The Meissner screening fraction of the $\sim$10~K superconductivity of this sample is still seen to be relatively small, of the order of $10$\%.  A slight Curie tail is observed in the highly-doped La samples at low temperatures, similar to that observed in \LaCa\ samples \cite{Saha1}, which may arise from paramagnetism associated with the FeAs lattice.

%%%%Figure Phase Diagram
\begin{figure}[tbh]
\begin{center}
\includegraphics[width=8.0cm]{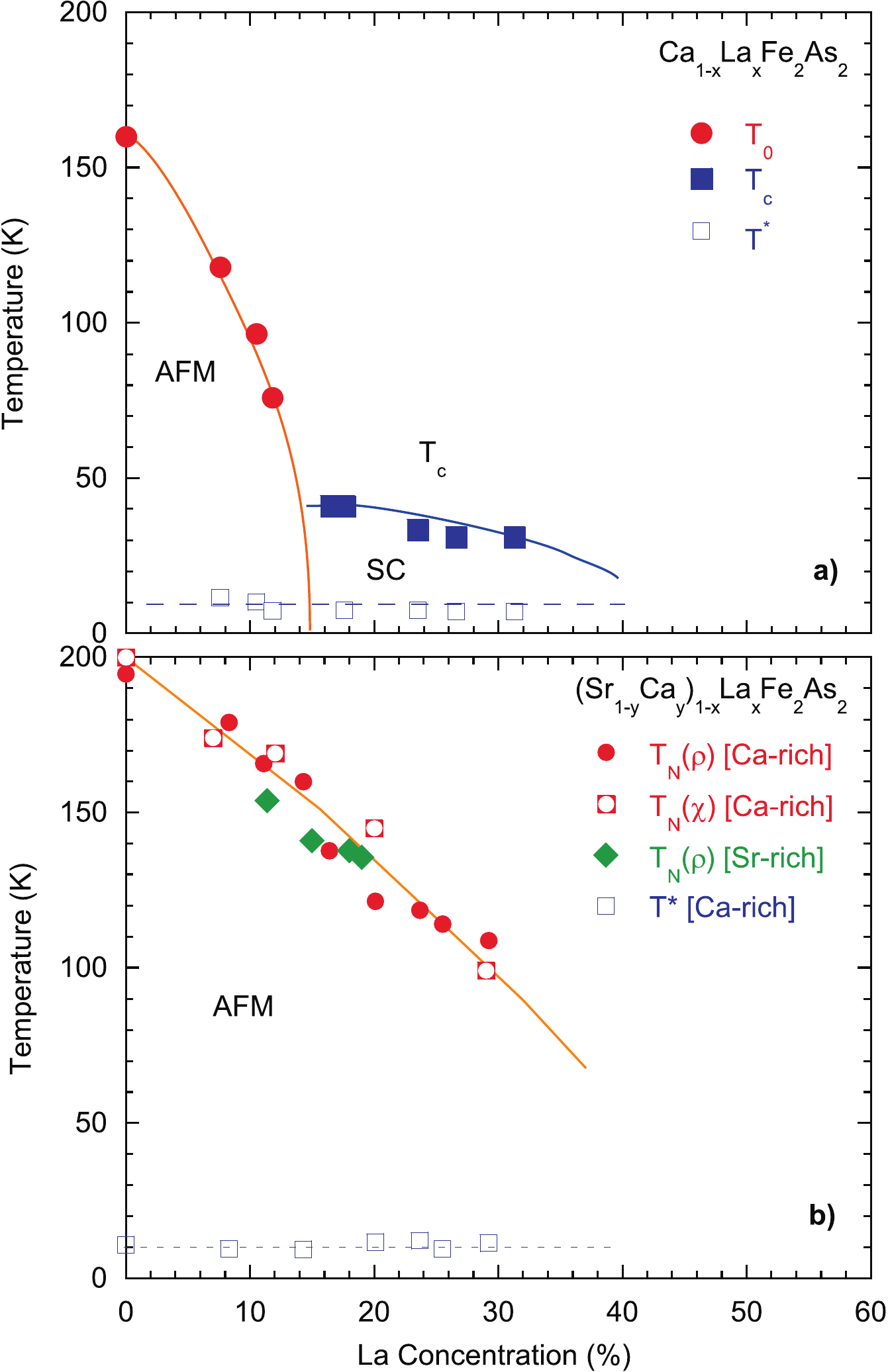}
\end{center}
\caption{\label{fig5}(a) The phase diagram for the \LaCa ~system~\cite{Saha1}, where high \tc~ superconductivity is induced on the border of AFM order and coexists with the $T^*\sim$10~K superconducting phase. (b)  Suggested phase diagram for the \SrCaLa ~system. The solid and broken lines are guides to the eye.}
\end{figure}
%%%%%%%%%%%%%%

Figure 5 presents a proposed phase diagram for the \SrCaLa ~system in comparison with that of the \LaCa ~system~\cite{Saha1}. As shown in FIG. 5b, a key observation in \SrCaLa ~is the absence of the high \tc ~ superconducting phase, which is observed ubiquitously in the rare-earth doped \CaF ~materials upon suppression of the AFM phase, despite a similar electron doping scheme. The lack of a high-\tc\ superconducting phase in \SrCaLa\ samples with La concentrations more than sufficient to induce superconductivity in \LaCa\ (FIG. 5a) suggests that a scenario where the superconducting phase arises solely from the presence of sufficient rare earth atoms that presumably cluster or percolate in some manner is improbable. The persistence of AFM order up to high concentrations of La may play a role here, as it seems as though high \tc ~superconductivity competes with AFM order and does not emerge until the complete suppression of magnetic ordering. Indeed, in every other rare-earth substituted $122$ system, $30$+~K superconductivity and antiferromagnetism are never found to coexist~\cite{Saha1}. This agrees with the occurrence of the highest \tc\ in the 1111 iron-pnictide family~\cite{Ren1,Paglione}, indicating that the highest \tc\ superconducting phase and magnetic ordering may be mutually exclusive.  Of course, further investigation will be necessary to bear out such a result. However, the conspicuous absence here of the high \tc ~phase, which has been thought to be an impurity phase of ReOFeAs, despite similar growth techniques and materials, lends credence to the idea that it is in fact intrinsic to the rare-earth substituted \CaF ~system.

\section{Summary}
In summary, we have studied the effect of electron doping by La substitution on \SrCaLa ~solid solutions by growing single crystals. We have constructed a phase diagram based on transport, magnetic susceptibility and structural characterization.  Chemical analysis indicates a strong inverse correlation between Sr and La.  Nonetheless, independent tunability of the $a$- and $c$-axis lattice parameters can be achieved.  The Sr-rich and Ca-rich regions show differing behavior;  in Ca-rich samples, antiferromagnetic ordering is found to coexist with superconductivity with $T^*\sim$10~K with a volume fraction $\sim10$\%. But, in contrast to \CaF, in neither case is \tn~fully suppressed and no high-\tc~ superconducting phase is observed, placing the constraint that complete suppression of AFM order is a necessary condition for the latter phase, which may provide an important clue for the superconducting pairing in the new iron-superconductors.

%
%
%%%%%%%%%%%%%%%% ACKNOWLEDGEMENTS %%%%%%%%%%%%%%%%%%
\vskip 0.5cm
\begin{center}
ACKNOWLEDGEMENTS
\end{center}
This work was supported by AFOSR-MURI Grant FA9550-09-1-0603.

%%%%%%%%%%%%%%%%%% BIBLIOGRAPHY %%%%%%%%%%%%%%%%%%%%

\end{document}